\documentstyle[12pt]{article}

\begin{document}
%%%%%%%%%%%%%%%%%%%%%%%%%%%%%%%%%%%%%%%%%%%%%%%%%%%%%%%%%%%%%%%%%%%%%%
\begin{flushright}
{\footnotesize\sf Portsmouth University\\
Relativity and Cosmology Group\\
{\em Preprint} RCG 96/04}
\end{flushright}
\[ \]
%%%%%%%%%%%%%%%%%%%%%%%%%%%%%%%%%%%%%%%%%%%%%%%%%%%%%%%%%%%%%%%%%%%%%%
{\Large\bf Anisotropic universes with conformal motion}
\[ \]
\[ \]
Roy Maartens$^{\dag}$ and Conrad M. Mellin$^{\ddag}$
\[ \] 
$^{\dag}$School of Mathematical Studies, Portsmouth University, 
Portsmouth PO1 2EG, England\\
$^{\ddag}$Department of Mathematics and Applied Mathematics, 
University of Cape Town, Cape Town 7700, South Africa.
\[ \]
{\bf Abstract.}   
By imposing natural geometrical and kinematical conditions on a 
conformal Killing vector in Bianchi I spacetime, we show that a class 
of axisymmetric metrics admits a conformal motion. This class 
contains new exact solutions of Einstein's equations, including 
anisotropic radiation universes that isotropise at late times.
\[ \]
PACS numbers: 0420J, 0440N, 9880H
\[ \]
\section*{1. Introduction}

A conformal Killing vector ({\sc ckv}) $\xi^\mu$
generates a local $G_1$ of
conformal motions of the metric:
\begin{eqnarray}
& & ds^2\equiv g_{\mu\nu}dx^\mu dx^\nu \quad\rightarrow
\quad e^\Phi ds^2 \quad\Leftrightarrow  \nonumber\\
& & {\cal L}_\xi g_{\mu\nu}\equiv \xi^\alpha\partial_\alpha g_{\mu\nu}
+g_{\alpha\nu}\partial_\mu\xi^\alpha
+g_{\mu\alpha}\partial_\nu\xi^\alpha =2\psi g_{\mu\nu}\label{1}
\end{eqnarray}
where $\psi$ is the conformal factor, with $\psi=0$ if
$\xi^\mu$ is a Killing vector ({\sc kv}). Conformally flat spacetimes
admit a maximal $G_{15}$ of conformal motions, with up to 10 of the
generators being {\sc kv}. Non--conformally flat spacetimes admit a
maximal $G_7$ of conformal motions, with at least 1 of the generators
being a proper {\sc ckv} (i.e. $\psi\neq 0$) \cite{hs}.

Solving the conformal Killing equation (\ref{1}) to find whether,
and under what conditions, a spacetime admits conformal 
motions, can lead to new solutions of the field equations
(see for example \cite{h}, \cite{mm}, \cite{mm2}, \cite{ct},
\cite{ct2}), 
or to new geometrical and kinematical insights (see for example
\cite{mmt}, \cite{mm3}, \cite{ct3}, \cite{mmtu}). As is well--known,
the standard Friedmann--Robertson--Walker ({\sc frw}) universes
are conformally flat, and admit 9 proper {\sc ckv}, all pointing
out of the homogeneous hypersurfaces \cite{mm3}. 

Spatially
homogeneous but anisotropic universes are not conformally flat,
and the existence of proper {\sc ckv} is subject to severe
constraints. These are the 
integrability conditions of (\ref{1}), including invariance
of the Weyl tensor:
\begin{equation}
{\cal L}_\xi C^\alpha{}_{\beta\mu\nu}=0
\label{2}
\end{equation}
Some conditions arise directly from (\ref{1}) via
kinematical considerations.
Let $u^\mu$ be the four--velocity field orthogonal to the homogeneous
hypersurfaces, so that its acceleration and vorticity are zero:
$\dot{u}_\mu=0=\omega_{\mu\nu}$. Then \cite{mmt}
\begin{eqnarray}
\psi &=& \left(u_\mu\xi^\mu\right)^{\displaystyle{\cdot}} \label{3}\\
{\cal L}_\xi u^\mu &=& -\psi u^\mu-h^{\mu\nu}\partial_\nu
\left(u_\alpha\xi^\alpha\right) \label{4}\\
0 &=& \sigma^{\mu\nu}\left[\sigma_{\mu\nu}+\left(h_{\mu\alpha}
\xi^\alpha\right)_{;\nu}\right] \label{5}
\end{eqnarray}
where $h_{\mu\nu}=g_{\mu\nu}+u_\mu u_\nu$ and $\sigma_{\mu\nu}$ is
the shear. (The first two equations hold in any spacetime without
acceleration and vorticity; the last holds in all spacetimes.)

It follows from (\ref{3}) that there are no proper {\sc ckv} tangent
to the homogeneous hypersurfaces, while (\ref{5}) shows that if
$\xi^\mu$ is parallel to $u^\mu$, then the shear vanishes and 
the spacetime degenerates to {\sc frw}. Thus any proper {\sc ckv} 
must be `tilted', neither parallel nor orthogonal to $u^\mu$.
It is natural to impose the condition that the conformal motion
of the metric should also be a conformal mapping of the 
four--velocity field $u^\mu$, i.e. that $\xi^\mu$ should map
$u^\mu$--curves into themselves \cite{mmt}:
\begin{equation}
{\cal L}_\xi u^\mu=-\psi u^\mu
\label{6}
\end{equation}
However, as shown in \cite{ct3}, there are no perfect fluid spatially
homogeneous anisotropic spacetimes with a proper {\sc ckv} that
satisfies the symmetry inheritance condition (\ref{6}).

Thus we are led to search for a proper {\sc ckv} that is tilted and
does not satisfy the simple inheriting condition (\ref{6}). The
simplest geometrical generalisation of (\ref{6}) is that $\xi^\mu$ 
be surface--forming with $u^\mu$. In section 2, we impose this
requirement, together with the requirement that $\xi^\mu$ forms
a Lie algebra with the {\sc kv} of the spacetime. Given
the complexity of the problem for general spatially homogeneous
spacetimes, we begin by tackling the simplest anisotropic 
generalisation of {\sc frw} spacetimes, the Bianchi I spacetimes.
Then the geometrical 
and kinematical requirements allow us to solve the conformal Killing
equation (\ref{1}). The integrability conditions force the metric
to be axisymmetric, and impose a second--order equation on the
metric components. 

For a perfect fluid, this means that
there is no freedom to specify an equation of state, 
and in section 3 we
show that all perfect fluid solutions which satisfy the weak energy
condition ($\rho\geq 0$) violate the dominant energy condition
($\rho+p \geq 0$). We consider fluids with anisotropic pressure, and
find new radiation solutions that 
isotropise at late times. This represents a possible
early--universe model with conformal motion.

\section*{2. Solutions of the conformal Killing equation}

The diagonal Bianchi I metrics have the form
\begin{equation}
ds^2=-dt^2+A_i(t)^2\left(dx^i\right)^2
\label{7}
\end{equation}
where $x^\mu=(x^0,x^i)=(t,\vec{x})$, and the canonical four--velocity
is $u^\mu=\delta^\mu{}_0$. They admit a $G_3$ of motions,
generated by the {\sc kv}
\begin{equation}
{\bf Y}_i\equiv Y_i{}^\mu\partial_\mu=\partial_i \quad {\rm with}
\quad [{\bf Y}_i,{\bf Y}_j]=0
\label{8}
\end{equation}
From the discussion in Section 1, we are led to look for a {\sc ckv}
$\xi^\mu$ 
that (a) closes with $\{{\bf Y}_i\}$ to
generate a $G_4$, and (b) is surface--forming with $u^\mu$:
\begin{eqnarray}
[{\bf Y}_i,\mbox{\boldmath $\xi$}]&=&
a_i\mbox{\boldmath $\xi$}+
b_i{}^j{\bf Y}_j  \label{9}\\
{\cal L}_\xi u^\mu &=&
-\psi u^\mu-\lambda h^\mu{}_\nu\xi^\nu \label{10}
\end{eqnarray}
where $a_i, b_i{}^j$ are constants, and $\lambda$ is some
scalar field, by Frobenius' theorem and (\ref{4}).

From (\ref{8}) and (\ref{9}) we get
\begin{equation}
\partial_i\xi^0=a_i\xi^0 \quad {\rm and}
\quad \partial_i\xi^j=a_i\xi^j+b_i{}^j
\label{11}
\end{equation}
The integrability condition of the second of equations (\ref{11}) is
$$
\partial_{[j}\partial_{k]}\xi^i=0\Rightarrow a_{[k}b_{j]}{}^i=0
\Rightarrow b_j{}^i=a_j b^i
$$
for some constant $b^i$ (square brackets denote
anti--symmetrisation). Thus (\ref{11}) integrate to give
\begin{equation}
\mbox{\boldmath $\xi$}=
e^{\vec{a}\cdot\vec{x}}\left[B(t)\partial_0+C^i(t)\partial_i\right]
\label{12}
\end{equation}
where we have used the freedom to add multiples
of the ${\bf Y}_i$ to $\mbox{\boldmath $\xi$}$
in order to set $b^i=0$.

Then (\ref{12}) and (\ref{10}) imply
\begin{eqnarray}
C^i(t) &=& C(t) c^i  \label{13}\\
\psi &=& \dot{B}e^{\vec{a}\cdot\vec{x}}  \label{14}\\
\lambda &=& {\dot{C}\over C}  \label{15}
\end{eqnarray}
where $c^i$ are constants. To avoid degenerate cases, we require 
$\dot{B}\dot{C}\neq 0$. Finally, the geometrically and kinematically
defined {\sc ckv} given by (\ref{12}) - (\ref{15}) is subject to the
conformal Killing equation (\ref{1}) for the metric (\ref{7}). It is
straightforward to show that the solution and integrability
conditions are:
\begin{eqnarray}
a_1B - c^1(A_1)^2\dot{C} &=& 0  \label {16}\\
{\dot{B}\over B}-a_1c^1{C\over B} &=& {\dot{A}_1\over A_1} 
\label{17}\\
A_2=A_3 &=& B  \label{18}\\
a_2=a_3 =c^2=c^3 &=& 0 \label{19}
\end{eqnarray}

It follows that such a {\sc ckv} can only occur in the axisymmetric 
class of Bianchi spacetimes, and that eliminating $C$
from (\ref{16}) and (\ref{17}) gives a condition involving only the
two independent metric functions. Writing $A\equiv A_1$ and
$a\equiv a_1$, we can summarise the results as:
\begin{eqnarray}
ds^2 &=& -dt^2+A^2dx^2+B^2(dy^2+dz^2)  \label{20}\\
\mbox{\boldmath $\xi$} &=&
e^{ax}\left[B\partial_t+a\left(\int {B\over A^2}dt\right)\partial_x
\right] \label{21}\\
\psi &=& \dot{B}e^{ax}  \label{22}\\
{a^2\over A^2} &=& {\ddot{B}\over B}-{\ddot{A}\over A}+{\dot{A}^2
\over A^2}-{\dot{A}\dot{B}\over AB} \label {23}
\end{eqnarray}
The integrability condition (\ref{23}) 
satisfies the conditions for `decomposability'
\cite{mn}, \cite{mw}, and thus may be reduced to
$$
A\left[A\left({B\over A}\right)^{\displaystyle{\cdot}}
\right]^{\displaystyle{\cdot}}=
a^2\left({B\over A}\right)
$$
This can be integrated upon defining a new time parameter:
\begin{eqnarray}
\tau &=& \int{dt\over A} \label{24} \\
B &=& \left(be^{a\tau}+ce^{-a\tau} \right)A \label{25}
\end{eqnarray}
where $b,c$ are constants. Clearly $a\neq 0$ to avoid the isotropic
{\sc frw} case ($B\propto A$).

\section*{3. Solutions of Einstein's field equations}

The class of metrics given by (\ref{20}), (\ref{24}) and (\ref{25})
is now subjected to Einstein's field equations 
\begin{equation}
R^\mu{}_\nu-{\textstyle{1\over2}}R\delta^\mu{}_\nu=
\rho u^\mu u_\nu+p h^\mu{}_\nu+\Pi^\mu{}_\nu   \label{26}
\end{equation}
where $\rho$ is the total energy density, $p$ is the isotropic
pressure, and $\Pi_{\mu\nu}$ is the trace--free anisotropic pressure
tensor. By symmetry there is no heat flux, and the anisotropic
pressure has the form
\begin{equation}
\Pi^\mu{}_\nu={\rm diag}\left(0,\Pi,-{\textstyle{1\over2}}\Pi,
-{\textstyle{1\over2}}\Pi\right)  \label{27}
\end{equation}
For a perfect fluid, $\Pi=0$, but $\Pi\neq 0$ in general, e.g.
for a perfect fluid with magnetic field (see \cite{ve}). The
distribution function for a kinetic gas may be expressed as \cite{mw2}
$$
f(x^\mu,p^\nu)=F(t,E)+F_\mu(t,E)e^\mu+F_{\mu\nu}(t,E)e^\mu e^\nu+
\cdots
$$
where $F_{\mu\cdots}$ are the covariant multipoles that determine
the anisotropy of the distribution,
and $p^\mu=Eu^\mu+\sqrt{E^2-m^2}e^\mu$, so that $E$ is the
particle energy and $e^\mu$ is a unit vector along the particle
3--momentum. The covariant dipole $F_\mu$ determines the heat flux 
(which in this case is zero), and the covariant quadrupole 
$F_{\mu\nu}$ determines the anisotropic pressure \cite{mw2}:
$$
\Pi_{\mu\nu}(t)={8\pi\over 15}\int_m^\infty \left(E^2-m^2\right)^{3/2}
F_{\mu\nu}(t,E)dE
$$
Thus $\Pi$ arises from a quadrupole anisotropy of the distribution
function, and by (\ref{27}):
$$
\Pi(t)={8\pi\over15}\int_m^\infty \left(E^2-m^2\right)^{3/2}
\Gamma(t,E)dE\quad{\rm where}\quad F^\mu{}_\nu={\rm diag}\left(
0,\Gamma, -{\textstyle{1\over2}}\Gamma,
-{\textstyle{1\over2}}\Gamma \right)
$$
The distribution cannot be isotropic unless the shear 
vanishes \cite{te}, so in general $\Pi\neq 0$ (see \cite{mm4},
\cite{r}, \cite{mes}, \cite{nm} for examples). For
collision--free and relaxational ({\sc bgk}) distributions,
the shear vanishes if the dipole, quadrupole and octopole are zero,
or if any 4 consecutive multipoles are zero, or if there are
only a finite number of non--zero multipoles \cite{etm}.

Whatever the physical source of anisotropic pressure is,
the field equations (\ref{26}) with (\ref{27}) and (\ref{20}) are
of the form
\begin{eqnarray}
{\dot{B}^2\over B^2}+2{\dot{A}\dot{B} \over AB} &=& \rho  \label{28}\\
-2{\ddot{B} \over B}-{\dot{B}^2\over B^2} &=& p+\Pi  \label{29}\\
-{\ddot{A}\over A}-{\ddot{B} \over B}-{\dot{A}\dot{B}\over AB} &=&
p-{\textstyle{1\over2}}\Pi  \label{30}
\end{eqnarray}
The system (\ref{28}) - (\ref{30}) is closed by (\ref{24}), (\ref{25})
and an equation of state when $\Pi\neq0$.

\subsection*{3.1 Perfect fluid solutions}

With $\Pi=0$, there is no freedom to specify an equation of state.
Subtracting (\ref{29}) from (\ref{30}) and using (\ref{24}) and
(\ref{25}), we get
\begin{eqnarray}
A^2 &=& k\left({1\over b^2e^{2a\tau}-c^2e^{-2a\tau}}\right) 
\label{31}\\
B^2 &=& k\left({be^{a\tau}+ce^{-a\tau} 
\over be^{a\tau}-ce^{-a\tau}}\right) \label{32}
\end{eqnarray}
where $k$ is a constant. Then (\ref{31}), (\ref{32}) 
and (\ref{28}), (\ref{29}) give
\begin{eqnarray}
\rho &=& {4a^2bc\over k}\left({b^2e^{2a\tau}+c^2e^{-2a\tau}+bc
\over b^2e^{2a\tau}-c^2e^{-2a\tau}}\right) \label{33}\\
p &=& -{4a^2bc\over k}\left({b^2e^{2a\tau}+c^2e^{-2a\tau}+3bc
\over b^2e^{2a\tau}-c^2e^{-2a\tau}}\right)  \label{34}
\end{eqnarray}
It follows from (\ref{33}) and (\ref{34}) that if $\rho\geq 0$,
then $\rho+p\leq 0$, so that the dominant energy condition is
violated, and the energy density
grows with expansion, by the energy conservation equation
$\dot{\rho}+(\rho+p)u^\mu{}_{;\mu}=0$.
The limiting case $\rho+p=0$
occurs if and only if $b$ or $c$ vanishes, which in turn
implies $\rho=0=p$ by (\ref{33}), (\ref{34}), and  $\psi=0$ by
(\ref{32}), (\ref{22}). In this case $\xi^\mu$ degenerates to
a null {\sc kv} of a vacuum Bianchi I solution. Taking $c=0$,
we have
\begin{eqnarray}
ds^2 &=& -dt^2+a^2(t-t_*)^2dx^2+{k\over b^2}(dy^2+dz^2) \label{35}\\
\mbox{\boldmath $\xi$} &=&
{\sqrt{k}\over b}e^{ax}\left[\partial_t+{1\over a(t_*-t)}\partial_x
\right] \label{36}
\end{eqnarray}
where $t_*$ is constant.

In summary: although perfect fluid solutions exist for {\sc ckv} that
are surface--forming with $u^\mu$, and do not exist at all for
`inheriting' {\sc ckv} (\ref{6})
that map $u^\mu$ conformally \cite{ct3}, 
these solutions are unphysical.

\subsection*{3.2 Anisotropic pressure solutions}

When $\Pi\neq0$, we impose the linear barotropic equation of state on
the isotropic pressure:
\begin{equation}
p=(\gamma-1)\rho
\label{37}
\end{equation}
where the constant $\gamma$ satisfies $1\leq\gamma\leq{4\over3}$
for `normal' fluids. By (\ref{28}) - (\ref{30}) and (\ref{37}),
and using (\ref{24}), (\ref{25}), we get an equation for $A(\tau)$:
\begin{eqnarray}
& &3{A'' \over A}+{\textstyle{3\over2}}(3\gamma -4){A'^2 \over A^2}
+2a(3\gamma -1)
\left[{be^{a\tau}-ce^{-a\tau}\over be^{a\tau}
+ce^{-a\tau}}\right]
{A'\over A}  \nonumber \\
& &+{\textstyle{1\over2}}a^2(3\gamma -2) 
\left[{be^{a\tau}-ce^{-a\tau}\over be^{a\tau}
+ce^{-a\tau}}\right]^2+2a^2=0 \label{38}
\end{eqnarray}
Thus we have reduced the problem to solving a single second order
(but non--autonomous) {\sc ode}. We will not consider this equation
in general, but only treat the autonomous case $c=0$, 
when (\ref{38}) becomes
\begin{equation}
3{A''\over A}+{\textstyle{3\over2}}(3\gamma -4){A'^2\over A^2}+
2a(3\gamma -1){A'\over A}+{\textstyle{1\over2}}a^2(3\gamma +2) =0
\label{39}
\end{equation}
Although the exact solution for 
$1\leq\gamma<{4\over3}$ has been found \cite{m}, 
we will only consider the more physically
interesting case of $\gamma={4\over3}$.
This is a radiation universe with anisotropy in the
distribution of radiation, and (\ref{39}) and (\ref{24}),
(\ref{25}) give
\begin{equation}
A=k(\tau-\tau_*)e^{-a\tau}\quad\quad B=kb(\tau-\tau_*)
\label{40}
\end{equation}
and
\begin{equation}
t-t_*={k\over a^2}\left[e^{-a\tau_*}-e^{-a\tau}\left\{
a(\tau-\tau_*)+1 \right\}\right]
\label{41}
\end{equation}
where $k, t_*, \tau_*$ are constants. Then (\ref{40}) with (\ref{28}),
(\ref{29}) implies
\begin{eqnarray}
\rho &=& {4e^{2a\tau}\left[3-2a(\tau-\tau_*)\right] \over
3k^2(\tau-\tau_*)^4}  \label{42}\\
\Pi &=& \left[{10a(\tau-\tau_*)+3 \over 2a(\tau-\tau_*)-3}\right]
{\rho\over 12}  \label{43}
\end{eqnarray}

\section*{4. Concluding remarks}

The exact solution (\ref{40}) - (\ref{43}) is an 
axisymmetric Bianchi I radiation
universe with anisotropy in the radiation distribution,
that admits a 
surface--forming conformal motion. 
There is a big bang at 
$\tau=\tau_*$, equivalently $t=t_*$, and for $a<0$ the universe
expands as $\tau$ and $t$ increase. The energy density and the
anisotropic pressure decay with expansion. The shear is given by
\begin{equation}
\sigma \propto {1\over A}\left| {A'\over A}-{B'\over B}\right|
\propto {e^{a\tau}\over \tau-\tau_*}
\label{44}
\end{equation}
and decays with expansion, so that the  solution isotropises
at late times. The anisotropic pressure starts from $-\infty$
at the big bang, increases through zero, 
at time $\tau=\tau_*-(3/10a)$, to a maximum, and then decays
like ${5\over 12}\rho$ at late times. 

Thus the solution has some reasonable physical
properties so that it might in principle be used as a model
of the early universe. However, we are not putting it forward as
such a model. Rather, we have shown how the imposition of natural 
geometrical and kinematical assumptions on a conformal motion
lead to solutions in Bianchi I spacetimes, including solutions
which are not immediately ruled out physically - as is often
the case with conformal motions.

The methods introduced here could be applied to other spacetimes,
including other Bianchi types. 
\[ \]

\end{document}